\documentstyle{article}
\title{
\bf  Potts Models on Feynman Diagrams 
}
\author{ {\it D.A. Johnston}\\
         Dept. of Mathematics\\
         Heriot-Watt University\\
         Riccarton\\
         Edinburgh, EH14 4AS, Scotland\\ \\
         and\\ \\
         {\it P. Plech\' a\v{c}}\\
         Mathematical Institute\\
         24-29 St Giles'\\
         Oxford\\
         OX1 3LB}
\date {28 April 1997}         
\begin{document}
  \maketitle
                      {\Large
                      \begin{abstract}
%
We investigate numerically and analytically Potts models
on ``thin'' random graphs -- generic Feynman diagrams, using
the idea that such models may be expressed as the $N \rightarrow 1$
limit of a matrix model.
The thin random graphs in this limit are locally tree-like, in distinction
to the ``fat'' random graphs that appear in the planar
Feynman diagram limit, 
$N \rightarrow \infty$, more familiar from discretized models
of two dimensional gravity.

The interest of the thin graphs is that they
give mean field theory behaviour for spin
models living on them without infinite range interactions or
the boundary problems of genuine tree-like structures
such as the Bethe lattice. 
$q$-state Potts models display a first order transition in the mean field
for $q>2$, so the thin graph Potts models provide a useful
test case for exploring discontinuous transitions in mean field theories
in which many quantities can be calculated explicitly in the saddle point
approximation. 

Such discontinuous transitions also appear in 
multiple Ising models on thin graphs and may have implications for the
use of the replica trick in spin glass models on random graphs.

%
                        \end{abstract} }
%
  \thispagestyle{empty}
%
%
  \newpage
%
                  \pagenumbering{arabic}

\section{Introduction and Reprise of Continuous Transitions}

A simple and elegant method of describing spin models on
random graphs, drawing inspiration from the matrix model methods \cite{0a}
used to describe
{\it planar} random graphs in two-dimensional gravity, was first
proposed in
\cite{0}. 
It was observed that the requisite ensemble of random graphs
of unrestricted topology
could be thought of as arising from
the perturbative Feynman diagram expansion of a scalar integral 
in much the same manner as the planar graphs 
that appear in two dimensional
gravity theories were generated from the perturbative expansion of a
matrix integral. In effect, the unrestricted random graphs 
appear in the $N \rightarrow 1$ limit of an $N \times N$
Hermitian matrix model, which
we denote as ``thin'' graphs, to distinguish them from
the planar ``fat'' graphs which 
appear in the $N \rightarrow \infty$ limit
and still retain their
matrix structure.
Throughout the paper we will use ``thin graphs'' and ``Feynman diagrams''
interchangeably to denote the random graphs
of unrestricted topology on which our spin models live.
Spin models on random graphs are of interest as
they will display mean field behaviour because
the graphs have a tree-like local structure \cite{0b}. The advantage
of using random graphs, which are closed, over genuine
tree-like structures such as the Bethe lattice is that dominant boundary effects
are absent. The complications, both analytical and numerical, of being forced
to consider only sites deep within the lattice are thus absent.
Other ways of accessing mean field behaviour, such as infinite range interactions,
are not very well suited for numerical simulation.

In previous papers we showed
that the thin graphs of the Feynman diagram expansion
offered a practical method
of investigating mean field models both analytically and numerically.
The equilibrium behaviour of ferromagnetic Ising models \cite{2} and spin glasses
\cite{3}
was found to parallel that of the equivalent model on the appropriate
Bethe lattice with the same number of neighbours
\cite{Bethe}, and the analytical treatment offered a
different perspective to previous approaches to random graph spin models
and spin glasses \cite{5,6}. 
The 
investigation of dynamical phenomena such as aging effects in
spin glasses \cite{4} was also found to be facilitated
by random graph simulations.
Other authors have also employed random
graphs in simulations of the random field Ising model \cite{7}
in order to avoid boundary problems with the Bethe lattice.    

Analytical calculations using the approach of \cite{0}
involve simple saddle point methods for standard
integrals, or quantum mechanical path integrals
in the case of continuous spins
\cite{8}. If we consider undecorated random graphs,
taking a $\phi^3$ theory for definiteness which will
generate 3-regular random graphs \footnote{We will
restrict ourselves to $\phi^3$ or 3-regular random graphs
throughout. The saddle point equations may still
be solved with larger numbers of neighbours, but become
rapidly more complicated.}, the 
number of such graphs with $2n$ vertices can be 
calculated as
\begin{equation}
N_n = {1 \over 2 \pi i} \oint { d \lambda \over
\lambda^{2n + 1}} \int_{-\infty}^{\infty} \exp \left( -\frac{1}{2} \phi^2 + \frac{\lambda}{6} \phi^3 \right)
\end{equation}
which, when evaluated using a saddle point approximation,
gives the correct counting 
\begin{equation}
N_n = \left( {1 \over 6} \right)^{2n} { ( 6 n - 1 ) !! \over ( 2 n ) !!
}.
\end{equation}
To include an Ising model we now decorate the 
vertices of the graphs
with Ising spins having a Hamiltonian
\begin{equation}
H = \beta \sum_{<ij>} ( \sigma_i \sigma_j - 1),
\end{equation}
where the sum is over nearest neighbour sites.
The partition function is then given by
\begin{equation}
Z_n(\beta) \times N_n = {1 \over 2 \pi i} \oint { d \lambda \over
\lambda^{2n + 1}} \int {d \phi_+ d \phi_- \over 2 \pi \sqrt{\det K}}
\exp (- S ),
\end{equation}
where $K$ is defined by
\begin{equation}
\begin{array}{cc} K_{ab}^{-1} = & \left(\begin{array}{cc}
1 & -c \\
-c & 1
\end{array} \right) \end{array}
\end{equation}
and the action itself is
a direct transcription of the 
matrix model action \cite{boul} to simple scalar variables
\begin{equation}
S = {1 \over 2 } \sum_{a,b}  \phi_a  K^{-1}_{ab} \phi_b  -
{\lambda \over 3} (\phi_+^3 + \phi_-^3).
\label{e3}
\end{equation}
The sum in the above runs over $\pm$ indices \footnote{We have
rescaled the $\phi$'s with respect to \cite{2,3,4}
for uniformity with the Potts model notation.}.
The coupling $c = \exp ( - 2 \beta )$ 
and the $\phi_+$ field can be thought of as representing ``up'' spins
with the $\phi_-$ field representing ``down'' spins.
It is necessary to include the counting factor $N_n$ to disentangle
the factorial growth of the undecorated graphs from any
non-analyticity due to phase transitions in the decorating spins.
One is also obliged to pick out the $2n$-th order in the expansion
explicitly with the contour integral over $\lambda$ as, unlike
the planar graphs of two dimensional gravity, 
$\lambda$ cannot be tuned to a critical value to cause a divergence. 

The mean field Ising transition manifests itself in this formalism
as an exchange of dominant saddle points. Solving the
saddle point equations $\partial S / \partial \phi_{\pm} = 0$
\begin{eqnarray}
\phi_+ &=& \phi_+^2 + c \phi_- \nonumber \\
\phi_- &=& \phi_-^2 + c \phi_+ 
\end{eqnarray}
(which we have rescaled to remove $\lambda $  and an irrelevant
overall factor) we find a symmetric high
temperature solution
\begin{equation}
\phi_+ = \phi_- = 1 - c
\end{equation}
which bifurcates at $c=1/3$ to the low temperature
solutions
\begin{eqnarray}
\phi_+ &=& { 1 + c + \sqrt{1- 2 c - 3 c^2} \over 2 } \nonumber \\ 
\phi_- &=& { 1 + c - \sqrt{1- 2 c - 3 c^2 } \over 2}.
\label{isingsol}
\end{eqnarray}
The bifurcation point is determined by the value of $c$ at which
the high and low temperature solutions for $\phi$ are identical,
which appears at the zero of the Hessian $\det ( \partial^2 S / \partial \phi^2 )$.
The magnetisation order parameter for the Ising model 
can also be transcribed directly from the matrix model \cite{boul} 
\begin{equation}
M = { \phi_+^3 - \phi_-^3 \over \phi_+^3 + \phi_-^3}
\end{equation}
and shows a continuous transition with mean field 
critical exponent ($\beta=1/2$). The other critical exponents may also be calculated
and take on mean field values.

Simulations of the Ising model are in very good agreement
with the analytical results \cite{2} even on a {\it single}
graph, which at first sight is rather surprising as the
saddle point calculations are formally for an annealed ensemble
of graphs. This appears to be true for all models where
one might expect self-averaging, such as ferromagnetic Ising
and Potts models -- one large graph can be thought of 
as a collection of smaller graphs in these cases. With spin glasses it is
still obligatory to consider a quenched ensemble of random
graphs in order to take a (quenched) average over the disorder.

\section{Potts Models}

The Hamiltonian for a q-state Potts model can be written 
\begin{equation}
H =   \beta \sum_{<ij>} ( \delta_{\sigma_i, \sigma_j} -1)
\end{equation}
where the spins $\sigma_i$ take on $q$ values.
The matrix model actions for such Potts models are well known
(though only solved exactly so far for $q=3$ \cite{Daul}).
For the 3-state Potts model the action is
\begin{equation}
S = { 1 \over 2 } ( \phi_1^2 + \phi_2^2 + \phi_3^2 ) - c ( \phi_1 \phi_2 + \phi_1 \phi_3 + \phi_2 \phi_3)
-{1 \over 3} ( \phi_1^3 + \phi_2^3 + \phi_3^3 ),
\label{potts3}
\end{equation}
which can be used as the action on thin graphs if one takes, as in the Ising case, the $\phi$'s to be
scalar variables.
For a q-state Potts model $c = 1/ ( \exp( 2 \beta)  + q-2)$.

\noindent
``Ising-like'' solutions to the 3 and 4-state Potts models were presented
in \cite{3}, the 3 state case being
\begin{eqnarray}
\phi_{1,2,3} &=& 1 - 2 c \; \; \;\;\;\;  \;\;\;\; \; \; \; \; \;   \; \; \; \; \;  \; \; \; \; \;\;\;\;\; \; \; \; \; \;   \; \; \; \;        (High T)  \nonumber \\
\nonumber \\
\phi_{1,2} &=& { 1 + \sqrt{1 - 4 c - 4 c^2} \over 2} \;\;\;\; \; \; \; \; \;   \; \; \; \; \;  \; \; \; \; \;  (Low T) ,\nonumber \\
\phi_3 &=& { 1 + 2 c - \sqrt{1 - 4 c - 4 c^2} \over 2}.
\label{potts3sol}
\end{eqnarray}
These high temperature and low temperature solutions are equal at
the zero of the Hessian
$c = 1/5$, i.e. $g=4$.
Similarly, the 4-state Potts model 
has the action
\begin{equation}
S = { 1 \over 2 } ( \phi_1^2 + \phi_2^2 + \phi_3^2 + \phi_4^2) - c ( \phi_1 \phi_2 + \phi_1 \phi_3 + 
\phi_1 \phi_4 + \phi_2 \phi_3 + \phi_2 \phi_4 + \phi_3 \phi_4)
-{1 \over 3} ( \phi_1^3 + \phi_2^3 + \phi_3^3  +\phi_4^3),
\label{potts4}
\end{equation}
and solving the saddle point
equations again 
gave Ising-like solutions
\begin{eqnarray}
\phi_{1,2,3,4} &=& 1 - 3 c \;\;\;\;\;\;  \;\;\;\; \; \; \; \; \;   \; \; \; \; \;  \; \; \; \; \;\;\;\;\; \; \;   
     (High T)  \nonumber \\
\nonumber \\
\phi_{1,2,3} &=& { 1 - c + \sqrt{1 - 6 c - 3 c^2} \over 2} \; \;  \; \;\;\;   (Low T), \nonumber \\
\phi_{4} &=& { 1 + 3c - \sqrt{1 - 6 c - 3 c^2} \over 2} 
\label{potts4sol}
\end{eqnarray}
where the solutions matched at $c=1/7$, $g=5$
\footnote{
The right hand side of the $Low T$ solutions may be exchanged,
as one might have expected on symmetry grounds, and we have used this freedom to put the solutions in a tidier form than in \cite{3}.}. 
The picture is repeated for higher q, where the action is
\begin{equation}
S = { 1 \over 2 } \sum_{i=1}^{q} \phi_i^2 - c \sum_{i<j} \phi_i \phi_j -{1 \over 3} \sum_{i=1}^q \phi_i^3
\label{qstate}
\end{equation}
and one finds a high temperature solution of the form $\phi_i= 1 - (q-1)c, \forall i$
bifurcating to a broken symmetry solution $\phi_i= \ldots \phi_{q-1} \ne \phi_q$
at $g=q+1$. 

These results are somewhat surprising
on two counts. Firstly,
the motivation for using thin graphs
was that they provided easy access to
mean field results. 
However, it is
known that the mean field theory for Potts models
predicts a {\it first} order transition for $q>2$.
Secondly, all the thin graph
results so far for various models have been identical
to the corresponding Bethe lattices, even down to non-universal
features
like the transition temperatures. 
Explicit calculations on the Bethe lattice have also
given first order behaviour for $q>2$ \cite{9}
and shown the values of $g$ obtained above for the 
3 and 4 state models correspond to spinodal points
on the Bethe lattice. The 
models hit a first order transition before attaining these
points. One might therefore expect that a first
order transition should be lurking in the saddle point
solutions for the actions above, given the previous tendency
for the thin graph results to slavishly parallel the
Bethe lattice.

The resolution of the conundrum is implicit already
in the solutions in equs.(\ref{potts3sol},\ref{potts4sol})
and their higher $q$ equivalents.
If we look at the Ising solution of equ.(\ref{isingsol})
we can see that the low temperature branches become
real exactly at the transition point, whereas the square roots
in the $q>2$ Potts solutions become real at larger c, and hence higher temperature.
The topology of the phase diagram is perhaps best understood by plotting
the magnetisation against c, which we do in Fig.1 for the Ising
solution and in Fig.2 for the 4 state Potts model
(all other $q>2$ state model s being of similar form).
Low temperature corresponds to small $c$ and high temperatures to
large $c$ for all the Ising and Potts models. 
For conformity with the Potts notation we define a
Potts style magnetisation for the Ising model as
\begin{equation}
m = { \phi_+^3 \over \left( \phi_+^3  + \phi_-^3\right)},  \; \;
{ \phi_-^3 \over \left( \phi_+^3  + \phi_-^3\right)}
\end{equation} 
where the first variant gives the upper (solid) branch
of the low temperature magnetisation curve in Fig.1 and the second
the lower (dashed) branch. Both definitions give the same value
for the horizontal (dotted) high temperature solution with $m=1/2$
as $\phi_+ = \phi_-$ there.
We can see that
the paramagnetic high temperature solution
bifurcates at {\bf O}
(c=$1/3$) to give the upper and lower magnetised low
temperature branches.

The equivalent magnetisation $m$ for the Potts models
is defined as
\begin{equation}
m = { \phi_q^3 \over \left( \sum_{i=1}^{q} \phi_i^3 \right)}
\end{equation}
on the {\it upper} low temperature branch
where it gives a maximum value
and
\begin{equation} 
m = { \phi_1^3 \over \left( \sum_{i=1}^{q} \phi_i^3 \right)}
\end{equation}
on the {\it lower} low temperature branch, where it
gives a minimum. On the symmetric high temperature branch (dotted)
both the definitions are, of course, equivalent
and one finds $m = 1/q$.
The standard Potts model order parameter can then be defined as
\begin{equation}
M = { q \max ( m ) - 1 \over q -1 }.
\end{equation}
which is zero in the high temperature paramagnetic phase and tends
to one in the magnetised low temperature phase.

The solutions
of equs.(\ref{potts4sol}) give
the lower (dashed) branch
in Fig.2,
descending from the point {\bf O} at which the square roots
become real. We can see that the magnetisation ``pitchfork'' of the Ising
diagram becomes skewed for $q=4$ (and all other $q>2$). The most important feature is that 
the upper (solid) branch does not connect continuously
with the (dotted) high temperature solution, which joins
the lower branch at {\bf P}. The upper branch is simply obtained for all
$q$ by choosing the opposite sign for the square roots to the lower branch.
As can be seen in Fig.2 it
corresponds to the true low temperature magnetised phase where $m \rightarrow 1$ 
(and hence $M \rightarrow 1$) as $T \rightarrow 0$. The first order transition,
denoted 
by a vertical line with ends labelled {\bf Q} in Fig.2
takes place when the free energy of the upper branch is equal to the
free energy of the high temperature solution. In the saddle point approximation
the free energy to lowest order in the number of vertices $n$ is just the action $S$ 
so the first order transition point is given by the $c$ value, and hence
temperature satisfying
\begin{equation}
S(upper \; branch)
= S( high \; temperature).
\label{PT}
\end{equation}
As one can see in Fig.2 as the temperature (i.e. $c$) is reduced a first order transition intervenes
between the {\bf Q}'s before {\bf P} is reached. Similarly, as the temperature is
increased from zero along the upper branch a first order transition occurs before {\bf O} is reached.
The portions of the magnetisation curve {\bf PO} and
the dotted horizontal line to the left of {\bf P} represent unstable states,
whereas {\bf PQ}, {\bf QO} and the lower dashed branch to the left of {\bf P}
represent meta-stable states \footnote{These considerations, and indeed
the two figures, are essentially identical to those for the Bethe lattice in \cite{9}.}. 
The meta-stable portions of the curve would be accessible by superheating
out of the magnetised phase ({\bf QO}) or supercooling from the paramagnetic
phase ({\bf QP}, {\bf P} $\rightarrow$ origin).
For completeness, we have listed the stable low temperature solutions for $q=3,4,5,6$
state Potts models in Appendix.A, which are the $q$ values simulated in the next section.

Various features of the solutions merit discussion. Equ.(\ref{PT})
can be solved analytically for moderate $q$
values on $\phi^3$ graphs without
too much difficulty, 
and all the solutions fit the
following
compact formula for the critical value of $c$ at {\bf Q}
\begin{equation}
c(Q) = { 1 - ( q - 1)^{-1/3} \over q -2}.
\label{ccrit}
\end{equation}
Indeed, if one takes the conjectured $q$ state solutions in Appendix.A
it is possible to write down the saddle point action
on the (upper) low temperature branch in terms
of $\phi=\phi_{1 \ldots q-1}$ and $\tilde \phi = \phi_q$
by substituting them into equ.(\ref{qstate})
\begin{eqnarray}
S = {1 \over 2} ( q - 1) \left[ 1 - c ( q - 2) \right] \phi^2  - { 1 \over 3} ( q -1) \phi^3
+ {1 \over 2} \tilde \phi^2  - {1 \over 3} \tilde \phi^3 - c ( q -1 ) \phi \tilde \phi
\label{app1}
\end{eqnarray}
and similarly on the high temperature branch
\begin{eqnarray}
S = { q \over 2} (1 - c ( q - 1) ) \phi_0^2 - { q \over 3} \phi_0^3
\label{app2}
\end{eqnarray}
where
$\phi_0 = 1 - ( q - 1) c$.
Setting (equ.(\ref{app1})) and 
(equ.(\ref{app2})) equal, one also obtains the same
$c(Q)$ as in equ.(\ref{ccrit}) above.

It is possible to calculate the jump in the magnetisation $\Delta M$
along the vertical line at {\bf Q} and one finds in all cases
\begin{equation}
\Delta M = { q - 2 \over  q -1}.
\label{mj}
\end{equation}
If one now refers back to the Bethe lattice calculations
of \cite{9} one can see that $\Delta M$ is identical to that observed on
the Bethe lattice. In addition, allowing for the differences in
conventions \footnote{The $\theta$
of \cite{9} is equal to $\exp ( - 2 \beta)$
in our notation.}, the formulae for the critical coupling $c(P)$ in equ.(22)
is also identical to that for the Bethe lattice transition.
The zero of the Hessian for the $q$ state Potts model action
gives us the value of $c$ at {\bf P} where the high temperature
solution joins the lower branch
\begin{equation}
c(P) = {1 \over 2 q - 1}
\end{equation}
and if we assume that the conjectured $q$ state low temperature solution
in Appendix.A is correct, we can also calculate the value
of $c$ at {\bf O} where the square roots become real
\begin{equation}
c(O) = { q -1 - 2 \sqrt{ q - 1} \over (q - 1 ) ( q - 5)}
\end{equation} 
($q=5$ can be handled by taking the limit $q \rightarrow 5$).
As $q$ is increased the separation
between points points {\bf O} and {\bf P} increases.
In Table.1 below we list for convenience the $c$ 
values
of points {\bf O}, {\bf P} and {\bf Q} for the $q=3,4,5,6$ state Potts models.

\begin{center}
\begin{tabular}{|c|c|c|c|} \hline
$q$   &  $c(O)$ & $c(P)$  & $c(Q)$              \\[.05in]
\hline
3 & 0.20711 & 0.20000 & 0.20630    \\[.05in]
\hline
4 & 0.15470  & 0.14286 &  0.15332    \\[.05in]
\hline
5 & 0.125 & 0.11111 & 0.12335   \\[.05in]
\hline
6 & 0.10557  & 0.09091 &  0.10380    \\[.05in]
\hline
\end{tabular}
\end{center}
\vspace{.1in}
\centerline{{\bf Table 1:} The $c$ values for the points {\bf O}, {\bf P}, {\bf Q} along with $\beta_{crit}$ for $q=3,4,5,6$ state Potts models} 

\medskip\noindent
The values of $c(O)$ and $c(P)$ that we have found
are again identical to those on the Bethe lattice.

\section{Simulations}

The acid test of the saddle point solutions is whether they match up
with simulations. We do not attempt a high accuracy 
verification of the analytical results here, but rather a consistency
check on the first order nature of the transition and a verification
of the values for $c(Q)$ and $\Delta M$ calculated in the previous section.
To this end we generated  single $\phi^3$ graphs with 250, 1000 and 2500
vertices  for each of the $q$ state models. We verified the results
by repeating all the simulations on a second, different graph for each size
with identical results within the error bars in all cases.

The generation of the graphs is easier than 
the planar $\phi^3$ graphs used in $2d$ gravity simulations because of the
absence of a constraint on the topology.
This  obviates the need to
use (for instance) Tutte's algorithm in producing the graphs. The simulation itself
used the Wolff algorithm, which will not beat the super-exponential
slowing down right at the first order transition point, but is expected
to be efficient elsewhere. We simulated a range of $\beta$ values,
allowing 20,000 equilibration sweeps followed by 20,000 $\times N$ cluster
updates, where $N$ was $O(10)$ - the exact value depending on the mean cluster size.
Measurements were made every $N$ cluster updates of all the standard
thermodynamic quantities, the energy $E$, the magnetisation $M$, specific heat $C$
and magnetic susceptibility $\chi$. We also measured various Binder's cumulants
for the magnetisation $<M^4>/<M^2>^2$ and energy $<E^4>/<E^2>^2$ as well
as correlation functions and autocorrelations.

We focus our attention first on the magnetisation curves for the various models,
which are presented, with fortuitous numbering, in Figs.3,4 for the $q=3,4$ state models
on graphs with 250,1000 and 2500 vertices. The smaller graphs display greater finite size rounding,
but by the time one has got to 2500 vertices the agreement with the magnetisation
calculated from the saddle point solutions (which are
formally for an infinite number of vertices) is already quite good . On both plots we have 
delineated the expected critical points and jumps in the magnetisations.
This agreement deteriorates somewhat for a given lattice size as $q$ 
and hence the strength of the transition increases. 
The $\beta_{crit}$ for various $q$ are estimated from the simulations by
looking at the crossing of Binder's magnetisation cumulant $<M^4> / <M^2>^2$ for the various graph
sizes. As one can see in Fig.5 for the 3 state Potts model (which is representative)
the errors in the measurement of the cumulant are quite large, but even given this 
the estimated critical temperatures are all close to those calculated in the
saddle point approximation. We also list the $\beta$ values for the 
two spinodal points {\bf O} and {\bf P} for comparison. As one can see
even the rather modest simulations carried out here are sufficient to show that
the spinodal point {\bf P} can be excluded as the transition point in all cases.
The results for all but the 3 state model cannot definitively exclude
the other spinodal point {\bf O} as the critical point, but the first
order nature of the transition and the value
of the jump in the magnetisation, as discussed below, favour a 
transition at {\bf Q} as predicted by the saddle point calculations.

\begin{center}
\begin{tabular}{|c|c|c|c|c|} \hline
$q$   &  3 & 4  & 5 & 6             \\[.05in]
\hline
$ \beta_{crit} $ & 0.674(2)   & 0.75(1)  & 0.81(1) & 0.87(1)    \\[.05in]
\hline
$ \beta(Q) $ & 0.67369   & 0.75451  & 0.81533 & 0.86441    \\[.05in]
\hline \hline
$ \beta(O) $ & 0.67122   & 0.74804  & 0.80472 & 0.84986    \\[.05in]
\hline
$ \beta(P) $ & 0.69315   & 0.80470  & 0.89588 & 0.97295    \\[.05in]
\hline
\end{tabular}
\end{center}
\vspace{.1in}
\centerline{{\bf Table 2:} The estimated $\beta_{crit}$, along with}
\centerline{the calculated $\beta(Q),\beta(O),\beta(P)$.}

\medskip

The
$\Delta M$ values are estimated by eye-balling
the magnetisation curves for largest graphs and are consequently to be taken
with a larger pinch of salt than the other measurements, but they are all
consistent with the $ (q - 2) / ( q -1)$ calculated in the previous section.
We tabulate the results for the magnetisation
jumps measured
from the simulations below in Table.3 along 
with $(q-2)/(q-1)$ for comparison.

\begin{center}   
\begin{tabular}{|c|c|c|c|c|} \hline
$q$   &  3 & 4  & 5 & 6             \\[.05in]
\hline   
$ \Delta M $ & 0.50(5)  & 0.68(1)  & 0.72(5) & 0.9(2)    \\[.05in]
\hline
$ (q-2)/ (q-1) $ & 0.5  & 0.66667  & 0.75 & 0.8    \\[.05in]
\hline
\end{tabular}
\end{center}
\vspace{.1in}
\centerline{{\bf Table 3:} The estimated $\Delta M$ along with}
\centerline{the calculated $(q-2)/(q-1)$.}

\medskip

Further confirmation that the transitions
are indeed first order can be obtained by looking at the values
of the Binder's energy cumulant near $\beta_{crit}$.
This is expected to scale to $2/3$ for a continuous transition and a
value less than $2/3$ if the transition is first order. 
The values
are listed in Table.4 and show a clear tendency to decrease around
$\beta_{crit}$ that grows stronger with increasing $q$.

\begin{center}
\begin{tabular}{|c|c|c|c|c|} \hline
$q$   &  3 & 4  & 5 & 6             \\[.05in]
\hline
${<E^4> \over <E^2>^2}$ & 0.664  & 0.65  & 0.64 & 0.60    \\[.05in]
\hline
\end{tabular}
\end{center}
\vspace{.1in}
\centerline{{\bf Table 4:} The values of Binder's energy cumulant at the estimated $\beta_{crit}$}

\medskip

The energy itself is calculable from
$E = - {\partial \log Z / \partial \beta}$ and is discontinuous at a first order transition. 
One finds values that are again in good agreement with the simulations by the time one has
reached 2500 vertices. The measurements for the 4 state Potts model are shown in Fig.6.
In a similar vein quantities such as the specific heat $C = \beta^2 ( {\partial^2 \log Z / \partial \beta^2})$
or the magnetic susceptibility may be calculated from the saddle point solutions and all give 
very satisfactory agreement
with the measured quantities in the simulations.  We do not describe these here as it is clear
from the results already presented that, even given the limitations of the fairly
modest simulations, there is ample support for the correctness of the saddle point solutions
and the picture of the first order transition that they suggest.

In closing, we note that something akin to a standard finite size scaling analysis is possible
with the thin graph approach to simulations, as witnessed by the use of the Binder's cumulant
to estimate the critical temperature in the current work. The place of the factor $L^{-1/\nu}$
that appears in finite size scaling on standard lattices, where $L$ is the linear size 
of the lattice, is taken by $n^{-1/\nu d}$ where $n$ is the number
of vertices in the graph. Although $d$ is formally infinite, the combination $\nu d$ is
still well defined and all the scaling relations may be written in terms of this.
Entirely analogous tactics have been used in the analysis of simulations
of spin models in planar diagrams in theories of two-dimensional gravity, where $d$ 
in this case was 
a dynamically generated fractal dimension that was {\it a priori}
unknown. On the analytical side $1/n$ corrections may be obtained (and have been obtained
already for the Ising model in \cite{2}) by calculating the determinantal corrections to the
saddle point solutions, so corrections to scaling can be obtained. Such issues would be
worth pursuing if very high accuracy verification of the correspondence between calculation
and simulation were required.

\section{Discussion}

Our previous analytical and numerical work \cite{2,3,4} on spin models
on Feynman diagrams had concentrated on the case
of continuous transitions. The results in the current paper show that
the simple saddle point equations that determine the phase structure
of such models are also adequate to describe first order
transitions.
In the continuous case the critical point was pinpointed by finding
the zeroes of the Hessian and corresponded to a bifurcation
of magnetised states from the unmagnetised high temperature solution.
This gives one the lower temperature spinodal point {\bf P} in the
Potts models, the true first order transition point at {\bf Q} being determined
by matching the saddle point actions (i.e. free energies) on the 
two branches of the solution. The upper spinodal point {\bf O} is fixed
as the point at which a square root appearing in the low temperature branches
becomes real.

The critical temperatures we have calculated, the jump in the magnetisation 
and the magnetisation curves themselves
are identical to the results obtained in \cite{9} on the Bethe lattice.
We thus conclude that the (mostly large) loops that are present in the
Feynman graphs have no effect of the critical behaviour of ferromagnetic
Potts models by comparison with the corresponding Bethe lattices. 
This is consistent with the earlier work on continuous transition which
also demonstrated Bethe-lattice-like results.
The loops will, however, have an effect in the antiferromagnetic models
considered in \cite{9}, where a two-step invariant measure
which presupposes a bipartite lattice was instrumental in the solution.
As already noted in \cite{2} for the antiferromagnetic Ising model,
loops of both even and odd length are present in the Feynman diagrams,
so frustration will be present on $\phi^z$ graphs if $q<z$. This
apparently leads to a spin glass phase rather than antiferromagnetic
order. In matrix models it is possible to arrange only even sided
polygonations by using complex rather than Hermitian matrices, but it is
not clear to us how to perform a similar trick on generic - rather than planar
- Feynman diagrams.

We find it rather remarkable that it is possible to write down the actions 
in equs.(\ref{app1},\ref{app2}) for 
both the high and low temperature branches for arbitrary $q$, based
on the ansatz for the $q$-state solutions in the Appendix.
This is on a par with the results in \cite{3} for the Hessian
of an arbitrary number of Ising replicas on thin graphs. The availability
of a solution where $q$ appears explicitly as a parameter opens
the possibility of exploring the $q \rightarrow 1$
limit of the model, related to percolation,
which we will address in a further publication. The percolative
transition on the Bethe lattice has some unusual features \cite{lmp},
which one might also expect to manifest themselves
on thin graphs. The behaviour of the model in an external field,
considered in some detail on the Bethe lattice in \cite{9}, can also be
investigated with the thin graph formalism both analytically and
numerically without much by way of complications over the results
presented here.

First order transitions also appear in the so-called Ising replica
magnet \cite{9a}, the finite $k$ version of the $k$-Ising replicas used
in taking the $k \rightarrow 0$ limit (``replica trick'') with quenched disorder.
In these models on thin graphs the spin glass transition temperature
appears as a sort of spinodal point \cite{3} for all $k>2$, with the transition
being continuous for $k=2$. 
A first order (in the overlap) transition intervenes before the 
putative continuous spin glass transition is reached for $k>2$. 
If it were not for this
one would be tempted to argue that the $k \rightarrow 0$ limit for the
spin glass transition temperature was trivial, as it is the same for {\it all} $k$.
The role of the first order transition
to a replica symmetric state when $k>2$ and the nature of the $k \rightarrow 0$ limit
for thin graph models thus requires further elucidation. It is possible
that the Potts models results presented here may cast some light
on its properties.

The numerical work in this paper was intended as a
consistency check of the formalism, rather than a full
scale numerical investigation and finite
size scaling analysis of the models. Nonetheless,
it is clear from the results presented that
the transitions {\it are} first order as predicted. The
agreement between the simulations and the saddle point calculations
for the critical temperatures and the observed jumps in the magnetisations
are very satisfactory even on graphs with 2500 vertices.

In summary, we have seen that the thin graph approach is a convenient
way of performing calculations and simulations for mean field Potts models
with first order phase transitions. Such models, as well as being of
interest in their own right, may help in understanding mean field spin glass
and percolative transitions.

\clearpage \newpage

\leftline{\Large \bf Appendix A}

\medskip

\noindent
We list here the low temperature solutions
(upper branch in Fig.2) 
for the $q=3,4,5,6$ state Potts models which are referred to in the text.
Note that the solutions appear follow a regular pattern 
and we have indicated the conjectured
$q$ state solution (which fits 
the cases listed below, the Ising model $q=2$,
and all the other higher $q$ solutions
we checked explicitly before exhaustion set in) at the end. Just as for the 
lower branch solutions where the signs in front of the square roots are reversed,
the right hand sides of the solutions can be exchanged
for a given $q$.

\bigskip

\leftline{\bf 3-State}
\begin{eqnarray}
\phi_{1,2} &=& { 1 - \sqrt{1 - 4 c - 4 c^2} \over 2} 
\nonumber \\
\phi_3 &=& { 1 + 2 c + \sqrt{1 - 4 c - 4 c^2} \over 2}.
\nonumber
\end{eqnarray}
\\
\leftline{\bf 4-State}
\begin{eqnarray}
\phi_{1,2,3} &=& { 1 - c - \sqrt{1 - 6 c - 3 c^2} \over 2}
\nonumber \\ 
\phi_{4} &=& { 1 + 3c + \sqrt{1 - 6 c - 3 c^2} \over 2}
\nonumber
\end{eqnarray}
\\
\leftline{\bf 5-State}
\begin{eqnarray}
\phi_{1,2,3,4} &=& { 1 - 2 c - \sqrt{ 1 - 8 c} \over 2} 
\nonumber \\
\phi_5 &=& { 1 + 4 c + \sqrt{ 1 - 8 c} \over 2}
\nonumber
\end{eqnarray}
\\
\leftline{\bf 6-State}
\begin{eqnarray}
\phi_{1,2,3,4,5} &=& { 1 - 3 c - \sqrt{ 5 c^2 - 10 c +1 } \over 2}
\nonumber \\ 
\phi_6 &=& { 1 + 5 c + \sqrt{ 5 c^2 - 10 c +1 } \over 2}
\nonumber
\end{eqnarray}
\\
\leftline{\bf q-State}
\begin{eqnarray}
\phi_{1 \ldots q-1} &=& { 1 -  (q-3) c - \sqrt{1 - 2 (q-1) c + (q-5) (q-1) c^2} \over 2}
\nonumber \\ 
\phi_q &=& { 1 +  (q-1) c + \sqrt{ 1 - 2 (q-1) c + (q-5) (q-1) c^2} \over 2}
\nonumber
\end{eqnarray}

\clearpage \newpage

\clearpage \newpage
\begin{figure}[htb]
\vskip 20.0truecm
\includegraphics{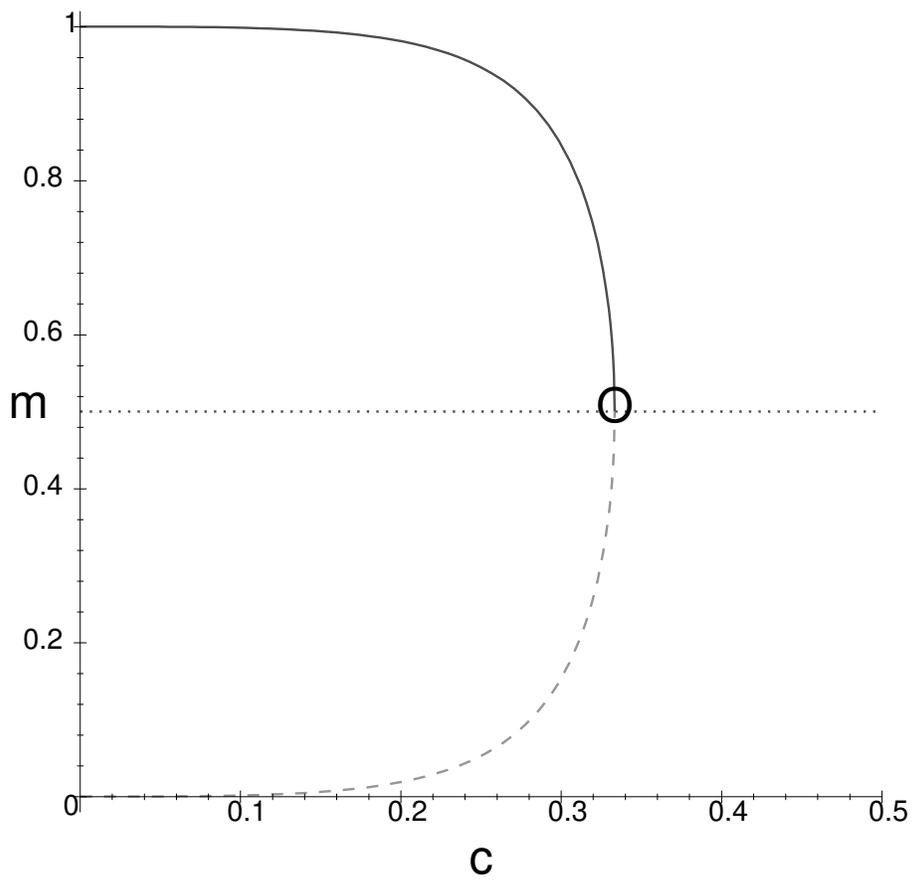}
\caption[]{\label{fig1} The magnetisation $m$
for the Ising model as calculated from
the saddle point solutions. The high temperature branch is
shown dotted, the upper low temperature branch solid and the lower
low temperature branch dashed.}
\end{figure}
\clearpage \newpage
\begin{figure}[htb]
\vskip 20.0truecm
\includegraphics{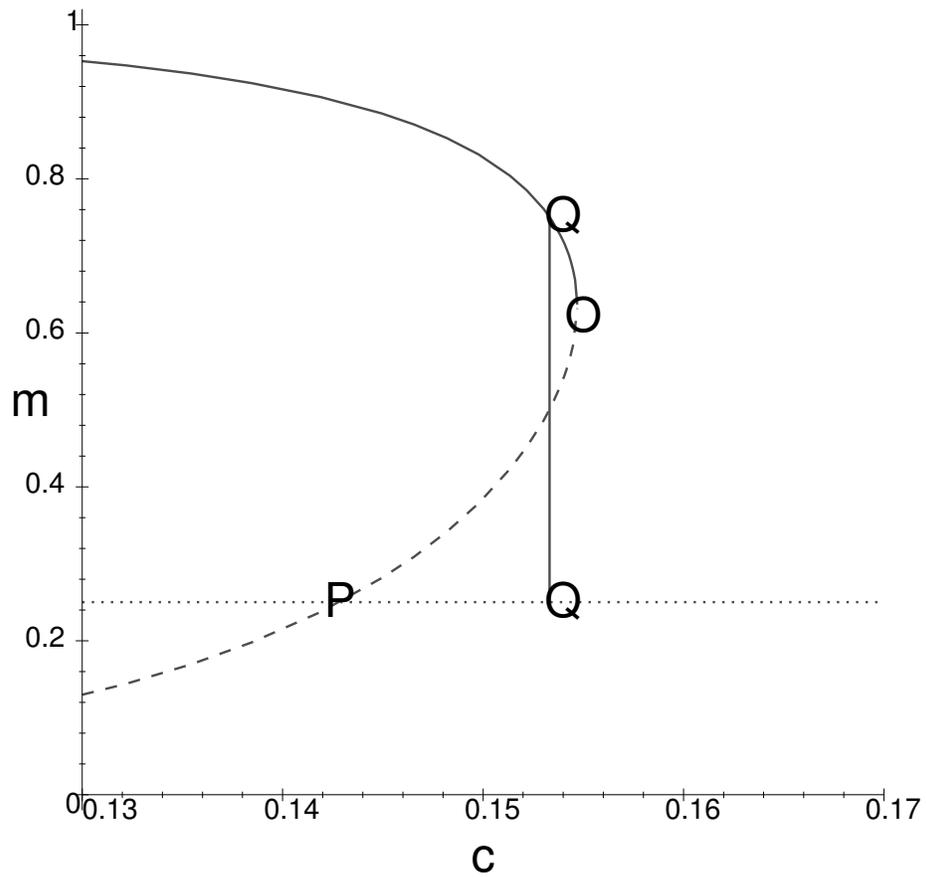}
\caption[]{\label{fig2} The magnetisation
$m$
for a $4$ state Potts model as calculated from
the saddle point solutions. The linestyles are as in Fig.1 and
only the portion of the graph close to the transition point is shown for clarity.}
\end{figure}
\clearpage \newpage
\begin{figure}[htb]
\vskip 20.0truecm
\includegraphics{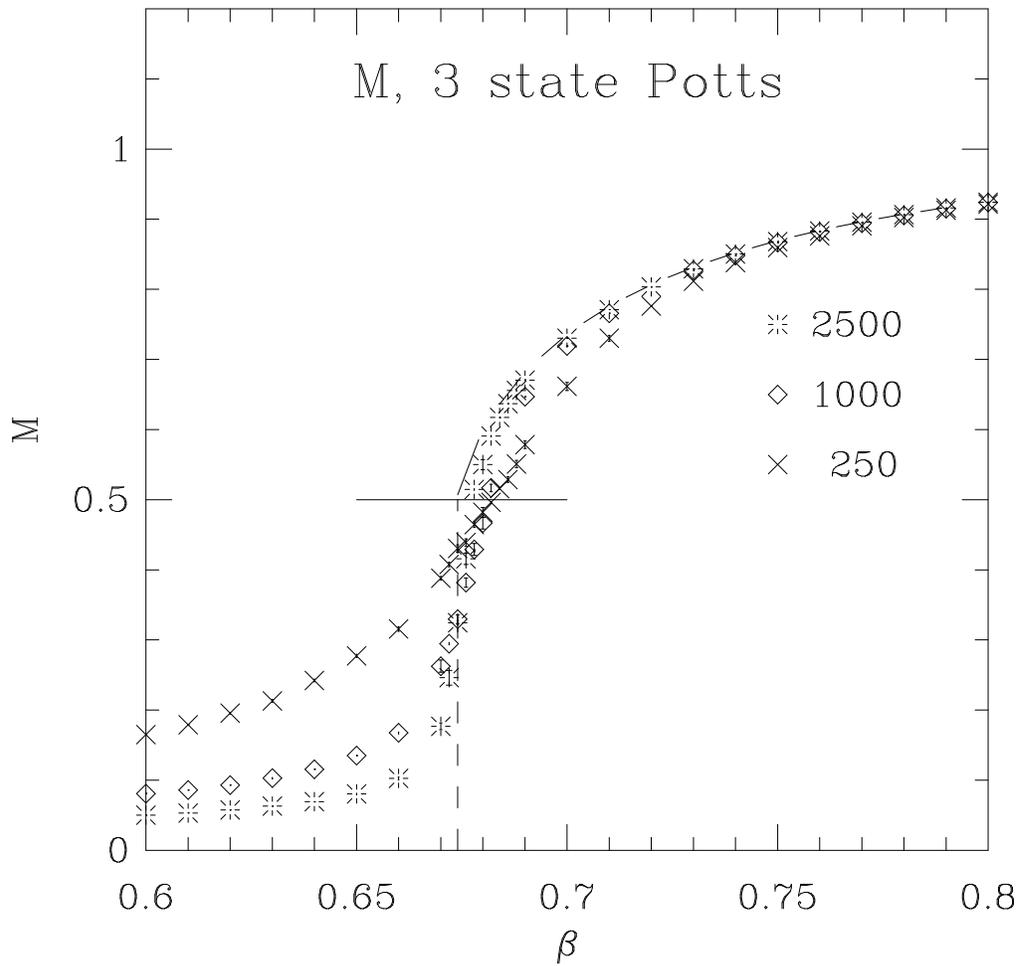}
\caption[]{\label{fig3} The magnetisation
$M$ ($= (q \max(m) -1)/(q-1)$)
for a $3$ state Potts model as measured in the simulations. The saddle point
solution is shown as a dashed line. The horizontal bar represents
the height of the ``jump'' $0 \rightarrow \Delta M$ in the magnetisation. 
The error bars are too small to be seen in all but the central points}
\end{figure}
\clearpage \newpage
\begin{figure}[htb]
\vskip 20.0truecm
\includegraphics{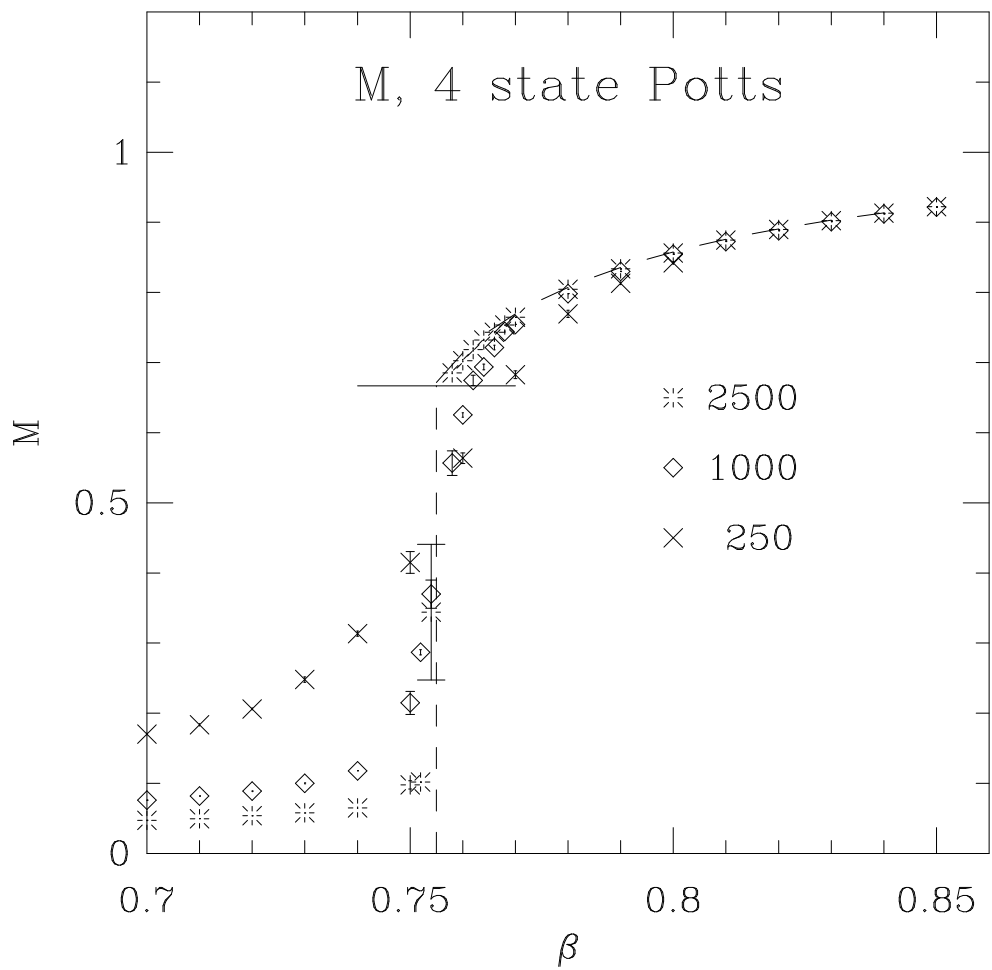}
\caption[]{\label{fig4} The magnetisation
$M$
for a $4$ state Potts model as measured in the simulations. Key as for Fig.3.}
\end{figure}
\begin{figure}[htb]
\vskip 20.0truecm
\includegraphics{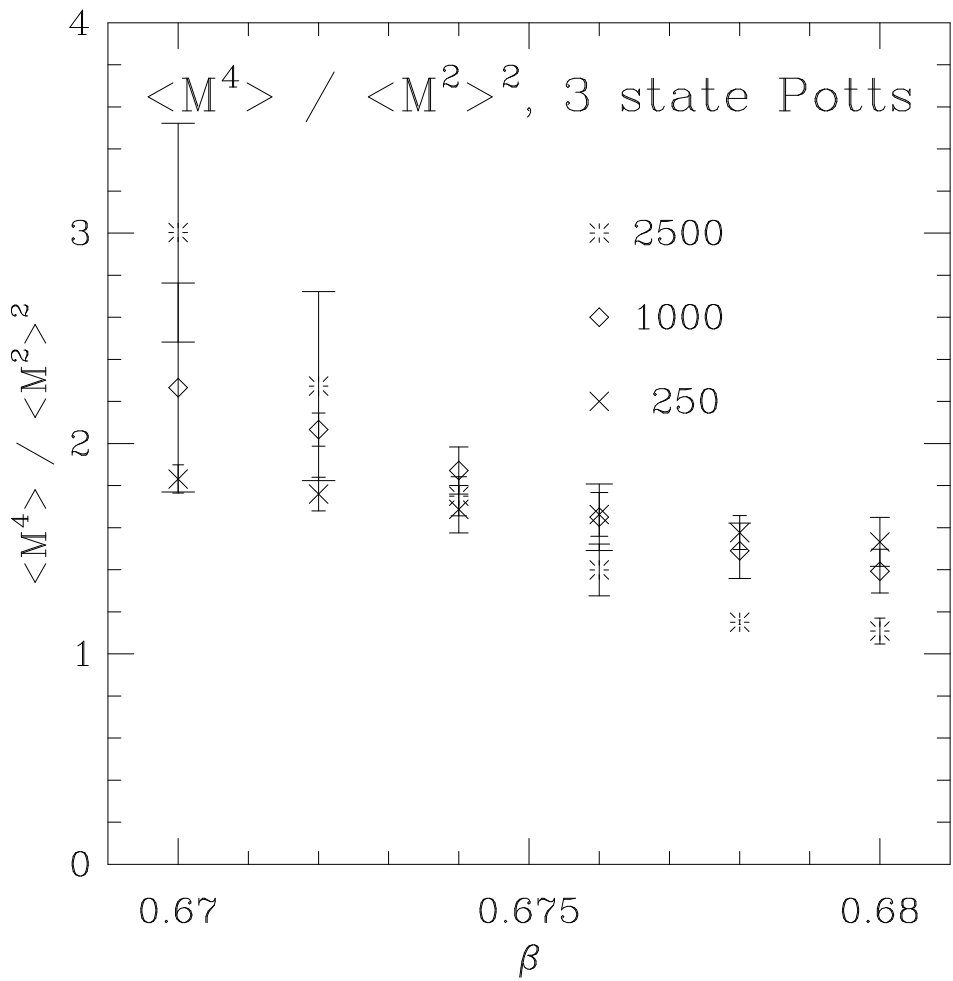}
\caption[]{\label{fig5} The crossing
of Binder's magnetisation cumulant 
for the $3$ state Potts model as measured in the simulations.}
\end{figure}
\clearpage \newpage
\begin{figure}[htb]
\vskip 20.0truecm
\includegraphics{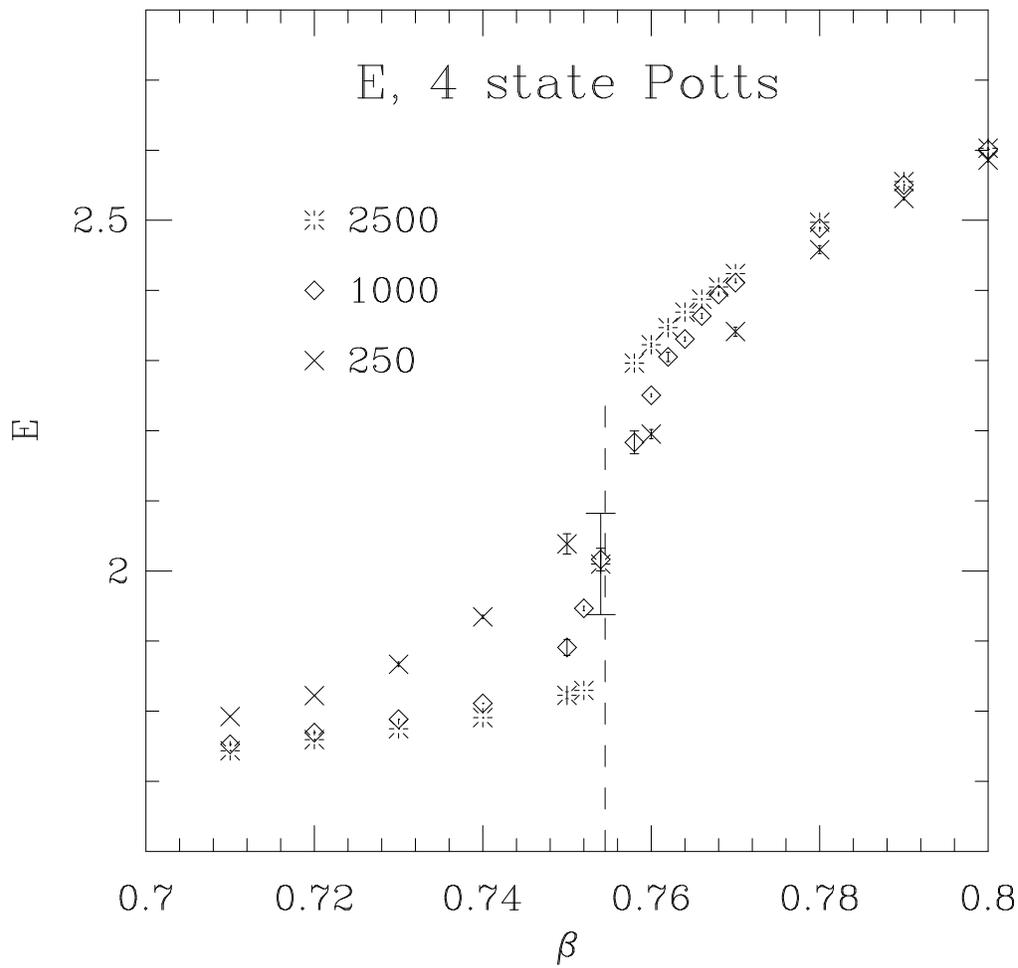}
\caption[]{\label{fig6} The energy
for a $4$ state Potts model as measured in the simulations.
The position of the transition
as calculated in the saddle point approximation is shown again as a dotted line.}
\end{figure}

\end{document}